\documentclass[aps,prb,twocolumn,pdfencoding=auto,superscriptaddress,longbibliography]{revtex4-2}
\usepackage[T1]{fontenc}
\usepackage[latin9]{inputenc}
\usepackage{amsmath,graphicx,amssymb}
\usepackage[pdftex, pdftitle={Article}, pdfauthor={Author},colorlinks,citecolor=blue,linkcolor=blue,urlcolor=blue]{hyperref}
\usepackage{amsfonts}
\usepackage{color}
\usepackage{chemformula}
\usepackage{cancel}
\newcommand{\bfk}{\mathbf{k}}
\newcommand{\bfq}{\mathbf{q}}
\newcommand{\phdag}{{\phantom{\dagger}}}

\usepackage{soul}
\usepackage{babel}

\makeatletter
\DeclareMathOperator{\Tr}{Tr}

\makeatother
\begin{document}
\title{Pair-Kondo effect: a mechanism for time-reversal breaking superconductivity  
in ${\rm UTe}_{2}$
}
\author{Tamaghna Hazra}
\thanks{tamaghna.hazra@kit.edu}
\affiliation{Center for Materials Theory, Rutgers University, Piscataway,
New Jersey 08854, USA}
\affiliation{Institut f\"ur Quanten Materialien und Technologien, Karlsruher Institut f\"ur Technologie, 76131 Karlsruhe, Germany.}
\affiliation{Institut f\"ur Theorie der Kondensierten Materie, Karlsruher Institut f\"ur Technologie, 76131 Karlsruhe, Germany.}

\author{Pavel A. Volkov}
\affiliation{Department of Physics, University of Connecticut, Storrs, Connecticut 06269, USA}
\affiliation{Department of Physics, Harvard University, Cambridge, Massachusetts 02138, USA}
\affiliation{Center for Materials Theory, Rutgers University, Piscataway, New Jersey 08854, USA}

\date{\today}

\begin{abstract}
An important open puzzle in the superconductivity of UTe$_2$ is the emergence of time-reversal broken superconductivity from a non-magnetic normal state. Breaking time-reversal symmetry in a single second-order superconducting transition requires the existence of two degenerate superconducting order parameters, which is not natural for orthorhombic UTe$_2$. Moreover, experiments under pressure [D. Braithwaite et. al., Commun. Phys. {\bf 2}, 1 (2019)] suggest that superconductivity sets in at a single transition temperature in a finite parameter window, in contrast to the splitting between the symmetry breaking temperatures expected for accidental degenerate orders. Motivated by these observations, we propose a mechanism for the emergence of time-reversal breaking superconductivity without accidental or symmetry-enforced order parameter degeneracies in systems close to a magnetic phase transition. We demonstrate using Landau theory that a cubic coupling between proximate magnetic order and magnetic moments of Cooper pairs (pair-Kondo coupling) can drive time-reversal symmetry breaking superconductivity that onsets in a single, weakly first order transition over an extended region of the phase diagram. We discuss the experimental signatures of such transition in thermodynamic and resonant ultrasound measurements. A microscopic origin of pair-Kondo coupling is identified as screening of magnetic moments by chiral Cooper pairs, built out of two non-degenerate order parameters - an extension of Kondo screening to unconventional pairs.
\end{abstract}
\maketitle

\section{Introduction}

The coexistence of superconductivity and 
magnetism in quantum materials brings with it a rich variety of possible unconventional superconducting phases, and many fundamental questions about their interplay~\cite{sigrist1991,annett1990,volovik1984,ueda1985,bulaevskii1985a,luke1998,joynt2002,yuan2003,dai2012,mineev2017,stolyarov2018,ribak2020,aoki2019,ran2019}. Here, we focus on one key question - how do Cooper pairs interact with local moments that are on the verge of ordering? 

\begin{figure}
    \centering
    \includegraphics[width=0.4\textwidth]{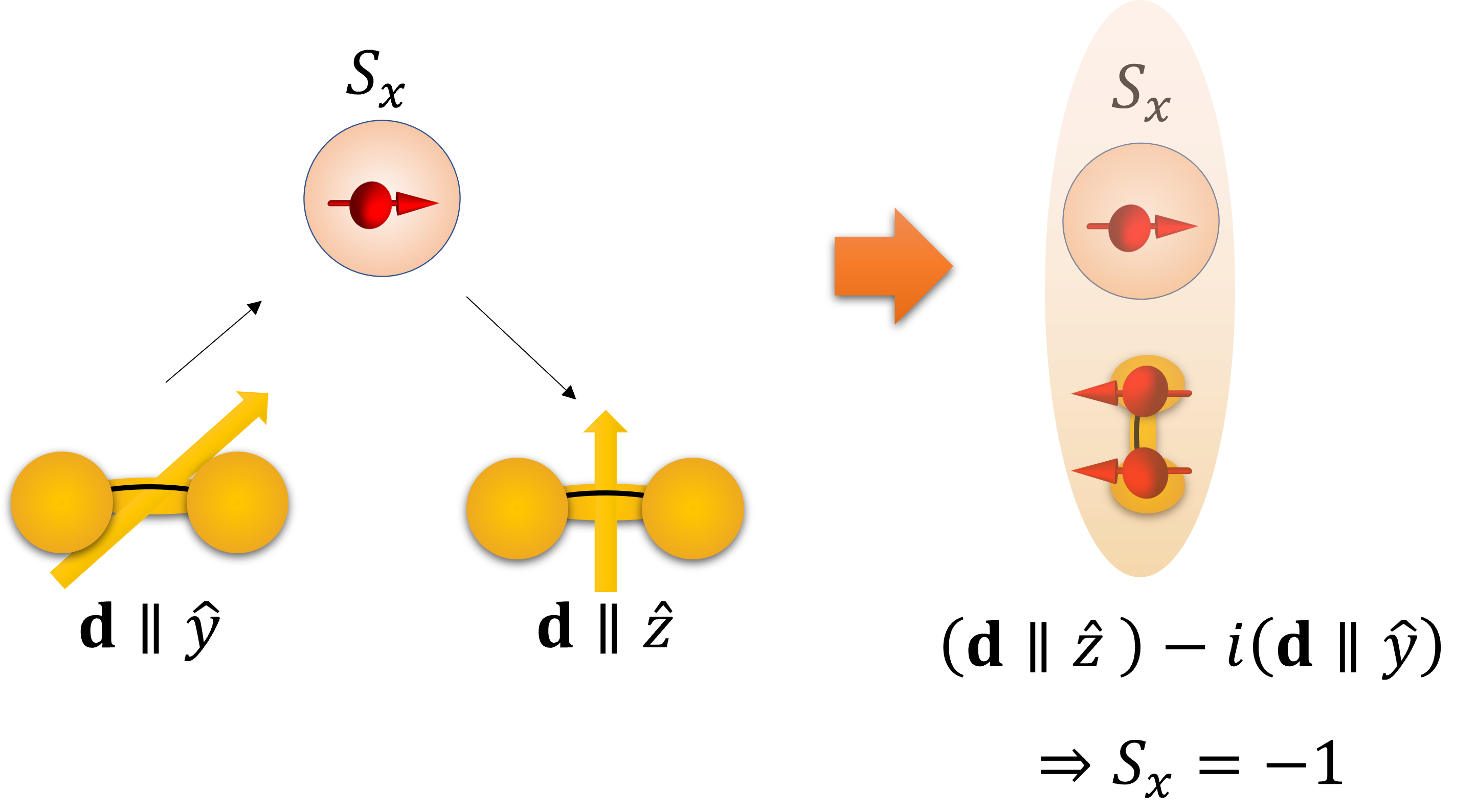}
    \caption{\textbf{Schematic demonstration of pair-Kondo coupling for triplet pairs:} An $S=1,S_y=0$ triplet pair scatters off a local moment into an $S=1,S_z=0$ state, rotating the d-vector about the $x$-axis along which the moment is aligned. The process is equivalent to an antiferromagnetic superexchange of the pair spin with the $S_x$ component of the moment.}
    \label{fig:schematic}
\end{figure}

For a conventional singlet superconductor, a single magnetic impurity can only be screened by Kondo coupling to the electron spin - so magnetic impurities are pair-breaking and host localized fermionic states inside the superconducting gap~\cite{yu1965,shiba1968,rusinov1969theory}. 
Pairs with internal degrees of freedom, such as spin or orbital, can however directly couple to the local moments (Fig.~\ref{fig:schematic}) and we present below a microscopic derivation of this exchange interaction, by which moments can be screened without pair-breaking.

A natural example of this local moment-superconductor interplay is provided by a number of heavy fermion superconductors~\cite{yuan2003,joynt2002,bulaevskii1985a,mineev2017,aoki2019} in which superconductivity emerges from a magnetic normal state, and coexists with magnetism at low temperatures. 
A more interesting class of materials involves the spontaneous breaking of time-reversal ($\mathcal{T}$) at a superconducting transition, a counter-intuitive phenomenon for two ordering tendencies that seem \emph{by definition} antagonistic. 
The most celebrated instance of this class is the A-phase of superfluid $^3$He~\cite{leggett1975, wheatley1975}, described by the equal-spin pairing potential of Anderson, Brinkman and Morel~\cite{anderson1961, anderson1973} $\Delta_{\uparrow\uparrow}=-\Delta_{\downarrow\downarrow}\sim (k_y+ik_z)/|\mathbf{k}|$ where $\mathbf{k}$ is the momentum. 
Another possible instance that has been extensively studied~\cite{mackenzie2003, mackenzie2017,pustogow2019} is ${\rm Sr_2RuO_4}$, supported by muon-spin relaxation~\cite{luke1998} and magneto-optical Kerr effect~\cite{xia2006} experiments both indicating the onset of $\mathcal{T}$ breaking at the superconducting transition temperature $T_c$, although the question of whether this is a \emph{bulk} time-reversal breaking transition is legitimately raised by the absence of any thermodynamic signature of a split transition under strain~\cite{hicks2014,li2021a,li2022b}.

From theoretical perspective, crystalline symmetry implies important restrictions on onset of time-reversal breaking superconductivity as a single transition.
In a lattice, spontaneous $\mathcal{T}$-breaking at the onset of superconductivity is usually expected only in high-symmetry crystals with at least two-dimensional irreducible representations (irreps). 
The superconducting order parameters belonging to non-degenerate one dimensional irreps, are simultaneously eigenstates of $\mathcal{T}$, as any complex phase can be absorbed into the $U(1)$ phase.
A condensate that breaks $\mathcal{T}$ must therefore contain a complex superposition of at least two such orthogonal pair eigenstates, either from the same~\cite{maiti2013} or distinct irrep. 
In the absence of two-dimensional irreps, $\mathcal{T}$-breaking superconductivity can then only arise from two distinct second order phase transitions with the two components of the complex pair wavefunction condensing at different temperatures.

A vexing challenge to this general symmetry argument is posed by the emergence of $\mathcal{T}$-broken superconductivity~\cite{hayes2021} from a non-magnetic normal state in UTe$_2$~\cite{aoki2022,devisser2019,coleman2022,nica2022a}, whose orthorhombic symmetry allows only one-dimensional irreps. 
While the samples in Ref.~\cite{hayes2021} did show signatures of a two-step transition, subsequent studies showed that when similar samples were cut up, the two transitions appeared to come from different fragments~\cite{thomas2021}. 
This suggests that intrinsically there is a single transition, with $T_c$ sensitively dependent on the crystal environment. 
Currently, samples with a single observable transition at ambient pressure have been grown reproducibly by various groups (see Fig. 1d of the original preprint \cite{sundar2022}) and the best available samples with the highest residual resistivity ratios (RRR) and $T_c$, consistently show no split in $T_c$~\cite{rosa2022,aoki2022a}. However, even in these samples, the splitting has been shown to appear under application of pressure above a critical value~\cite{braithwaite2019,aoki2020multiple}, with no phase boundary observed between the ambient pressure superconductor and the superconducting phase at high pressure below the split $T_c$s.

We note that the relation between splitting of $T_c$s at zero pressure and time-reversal breaking is actively under investigation. Among the samples with no split in $T_c$, there is a subset of crystals that are grown in a molten salt flux in which zero-field muon spin relaxation experiments are unable to discern any spontaneous magnetic fields in the superconducting state~\cite{azari2023b}, and a subset of crystals grown by chemical vapor transport in which muons do feel a static random magnetic field in the superconducting state, suggestive of moments frozen in a glass~\cite{sundar2023}. Moreover, local internal fields in these latter samples are also supported by local magnetic susceptibility measurements which can resolve spontaneous vortex-antivortex pairs in zero external field~\cite{iguchi2023}. A recent paper~\cite{ajeesh2023} has reported absence of a spontaneous Kerr effect in both types of samples with no split in $T_c$ (salt-flux and CVT grown). However, under pressure the new generation of salt-flux grown samples have been also been shown to elicit a split in $T_c$ beyond a finite critical pressure~\cite{wu2024}.

\begin{figure*}
    \centering
    \includegraphics[width=\textwidth]{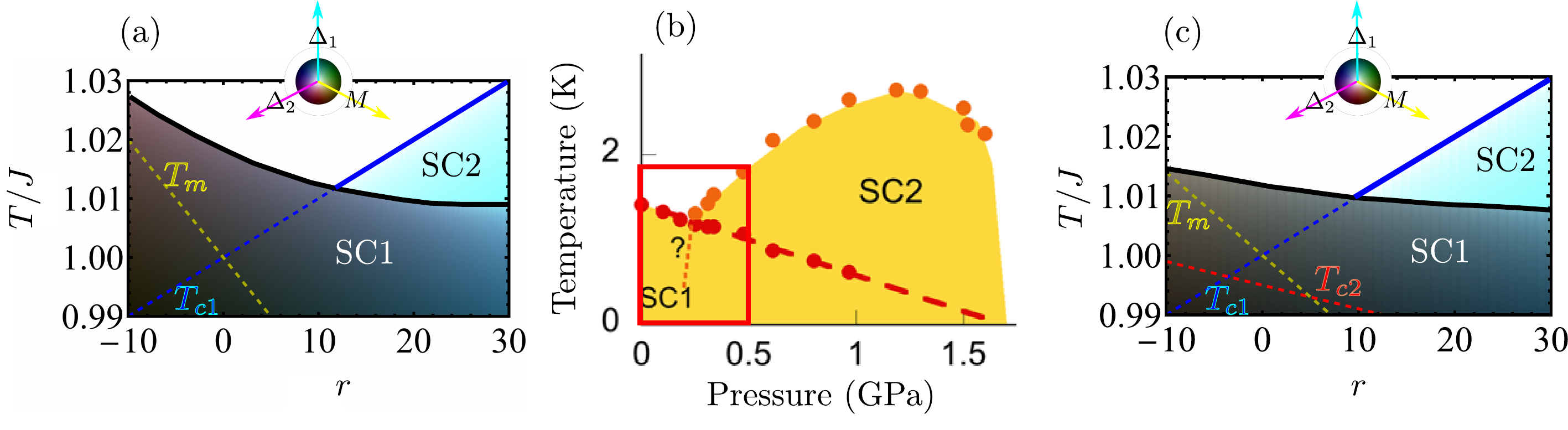}
    \caption{(a) Theoretical phase diagram as a function of temperature and tuning parameter $r$. Solid black (blue) line is a first(second)-order phase boundary. SC2 is a $\mathcal{T}$-invariant triplet superconductor with $\Delta_1\neq0,\Delta_2=M=0$. SC1 is a $\mathcal{T}$-broken superconductor with $\Delta_1,\Delta_2,M\neq0$. Dashed lines indicate the underlying critical temperatures in absence of coupling between magnetic and superconducting orders: $T_{c1}/J=1+0.001r, T_m/J=1-0.002r, a_2/J=0.005(1+0.04 r), b_1=b_2=b_M=2J,b_{1M}=b_{2M}=b_{12}=0,\alpha_1(1+0.001r)=\alpha_m(1-0.002r)=1$. Color scale is defined by the RGB values Red=$255(1-\Delta_1/{\rm Max})$, Green=$255(1-\Delta_2/{\rm Max})$, Blue=$255(1-M/{\rm Max})$ with ${\rm Max}=0.15J$. c.f. (b) Experimental phase diagram from AC calorimetry (from Ref.~\cite{braithwaite2019,aoki2020multiple}) as a function of pressure and temperature, where the second transition at low pressure (dashed) is not observed. The observed phase boundaries at low-pressure (boxed) are qualitatively captured by the free energy in \eqref{eq:FreeEn}. (c) Theoretical phase diagram for a different parameterization of the free energy in \eqref{eq:FreeEnalt} with $T_{c1}/J=1+0.001r, T_{c2}/J=0.995(1-0.0004r), T_m/J=1-0.0014r,b_1=b_2=b_M=2J=2J,b_{1M}=b_{2M}=b_{12}=0,\alpha_1(1+0.001r)=0.995(1-0.0004r)\alpha_2=\alpha_m(1-0.0014r)=1$}
    \label{fig:PhaseDiag}
\end{figure*}

The initial observations of potential time-reversal breaking in UTe$_2$ has motivated two scenarios: that $\mathcal{T}$-breaking at $T_c$ is present only in inhomogeneous samples~\cite{willa2021a} with split $T_c$ or there is an accidental near-degeneracy between two superconducting order parameters, with the split in $T_c$ being below experimental resolution in samples with high $T_c$ and RRR. 
In~\cite{shaffer2022}, a mechanism was proposed to stabilize such an accidental degeneracy by studying the renormalization group flow of interactions in an accidentally $C_4$ symmetric bandstructure. However, such an accidental degeneracy should split linearly in the presence of non-symmetry breaking perturbations~\cite{zinkl2021}, such as pressure. 
The vanishing of the splitting over an extended range of pressure, rather than a single fine-tuned pressure, presents a conundrum in the observed phase diagram in Fig. \ref{fig:PhaseDiag}(b) - two second-order phase boundaries cannot merge into one~\footnote{The phase diagram in \cite{braithwaite2019,aoki2020multiple} is constructed from the bulk specific heat on two different samples, making effects of inhomogeneity or impurity phase unlikely to explain it.}
This observation also poses a clear challenge for the scenario of single-component superconductivity at low pressures, including the ambient pressure systems. 
The possibility that we explore in this work is that the observed phase boundaries in Fig. \ref{fig:PhaseDiag}(b) are not all continuous second-order transitions.

Here we present a generic mechanism for a weakly first-order transition into a $\mathcal{T}$-broken superconductor in systems close to a magnetic instability.
The first-order character of the transition allows the simultaneous appearance of multiple order parameters at $T_c$, necessary to explain a single $\mathcal{T}$-breaking superconducting transition. As is shown below, it is sufficient to have a near degeneracy between one superconducting order and a magnetic order, rather than two superconducting order parameters.
There is extensive evidence that above the superconducting $T_c$ UTe$_2$ is also on the verge of a ferromagnetic instability, 
from a divergent $a$-axis magnetic susceptibility~\cite{ran2019}, 
a critical slowing down of magnetic fluctuations in muon-spin rotation~\cite{sundar2019}, 
emergence of low-frequency magnetic fluctuations in nuclear magnetic resonance~\cite{tokunaga2019} and 
a divergent ferromagnetic susceptibility seen in field trained polar Kerr effect measurements~\cite{wei2022}. 
If the moments fluctuate slowly on the timescale of electronic motion~\cite{machida2020theory}, even in inhomogeneous samples, puddles of magnetic order~\cite{sundar2023} can harbor puddles of magnetic superconductivity over the magnetic correlation length. 
When these inhomogenous puddles grow with decreasing temperature, there is a first-order transition into a state with phase coherence and long-range superconducting order over the entire sample.

\begin{figure}[h!]
    \centering
    \includegraphics[width=0.4\textwidth]{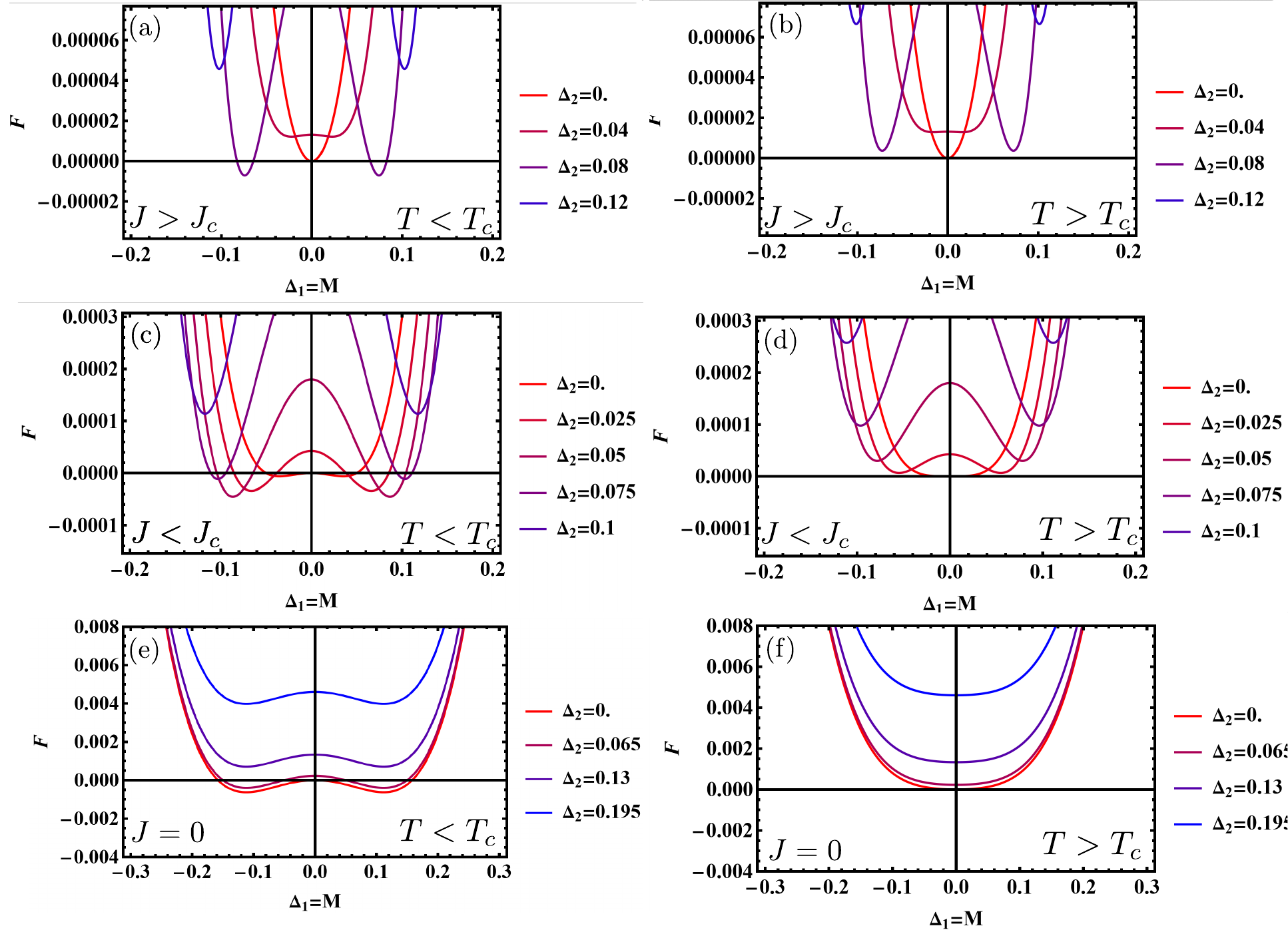}
    \caption{
    Free energy profile, Eq. \eqref{eq:FreeEn}, as a function of order parameter magnitudes for temperature $T$ above (a,c,e) and below (b,d,f) the onset of superconductivity. We normalize all quantities by setting $T_{c1}=T_m\equiv 1$; the other parameters are $b_1=b_2=b_M=2,b_{1M}=b_{2M}=b_{12}=0,\alpha_1=\alpha_m=1$. (a,b) For $J=1,a_2=0.005$, a first-order transition occurs between $T=1.018$ and $T=1.019$: in the ordered state~(a), the local minimum at $\Delta_1=M=0$ (red arrow) coexists with the global minimum at $\Delta_1=M=0.08$. (c,d) For $J=1,a_2=0.067$, a second-order transition occurs at $T=1$. In the ordered state~(c), there is no minimum at $\Delta_1=M=0$. For $T>1$ (d), the only true local minimum is at $\Delta_1=M=\Delta_2=0$. The apparent minima as a function of $\Delta_1=M$ are unstable to decreasing $\Delta_2$, which smoothly connects them to the $\Delta_1=M=\Delta_2=0$, forming two "valleys". (e,f) For $J=0,a_2=0.005$, a second-order transition occurs at $T=1$ and neither additional local minima nor valleys are present for $T>1$.
    }
    \label{fig:FreeEn}
\end{figure}

To capture this effect in a homogeneous system, we introduce an interaction of unconventional pairs with local moments - an extension of Kondo coupling to electron pairs with an internal degree of freedom, such as spin. 
This is captured by the local Hamiltonian
\begin{equation}
H_{PK}=
-iJM_{\Gamma\times\Gamma^{'}}\left(
b_{\Gamma^{\phantom{'}}}^{\dagger}b_{\Gamma^{'}}^\phdag -b_{\Gamma^{'}}^{\dagger}b_{\Gamma^{\phantom{'}}}^\phdag
\right)\label{eq:PairKondo}
\end{equation}
where \textbf{$b_{\Gamma(\Gamma^{'})}^{\dagger}$ }creates a pair in the one-dimensional irrep $\Gamma(\Gamma^{'})$ and the pair bi-linear couples to the component of the local moment that transforms under the product of the two irreps. 
This provides a generic route for coupling unconventional pairs to proximate magnetism, so that unconventional pairs can gain energy by coupling to slowly-fluctuating moments as a chiral combination $b_{\pm}=(b_{\Gamma^{\phantom{'}}}\pm i b_{\Gamma^{'}})/\sqrt{2}$ of the pairs $b_{\Gamma(\Gamma^{'})}^{\dagger}$ which are eigenstates of $\mathcal{T}$ and point group symmetries.
As a result, even when only one pair component is near critical, coupling to proximate magnetism can induce the second component, forming local $\mathcal{T}$-breaking pairs that can condense above the underlying superconducting and magnetic critical temperatures. 
Note that this energy gain is available as long as the fluctuations of the moments are slow compared to the pair kinetic energy,  whether the moments are near-critical, as we consider here or frozen in a glass, as suggested by Ref.~\cite{sundar2023}.

\begin{figure}
    \centering
    \includegraphics[width=0.45\textwidth]{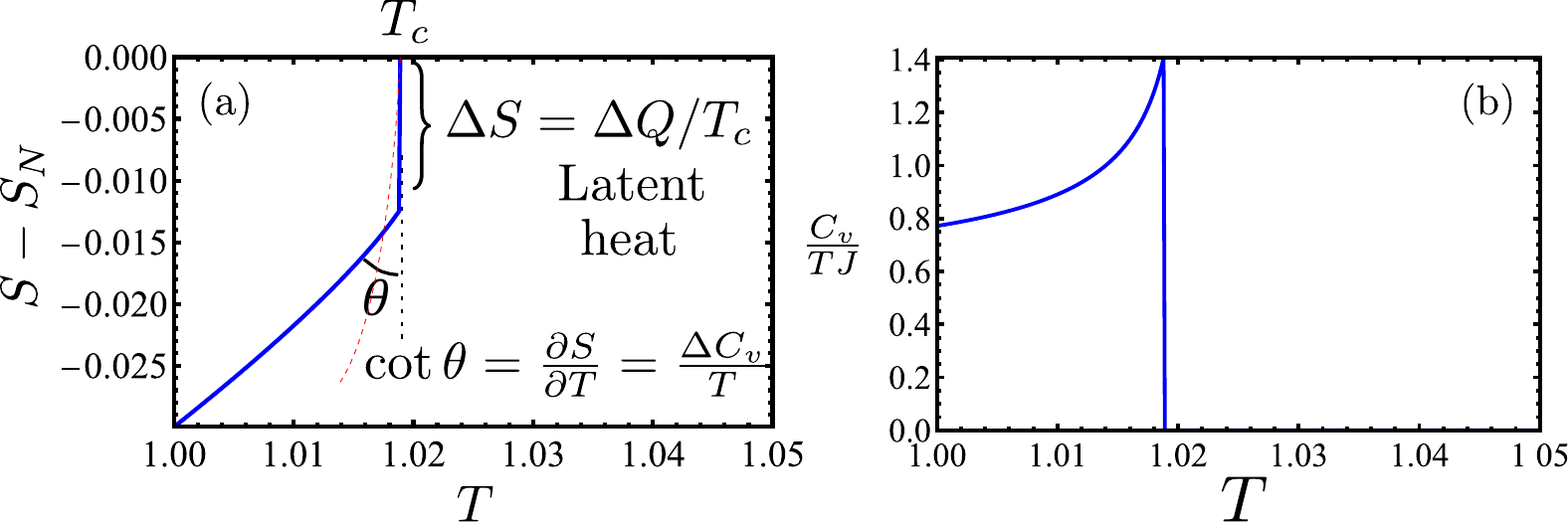}
    \caption{(a) Entropy (blue) and (b) specific heat across the first-order transition for $\alpha_1=\alpha_m=T_{c1}=T_m=J,b_1=b_2=b_M=2J,b_{1M}=b_{2M}=b_{12}=0,a_2=0.005J$ c.f. experimentally measured specific heat in Ref.~\cite{aoki2022a}. As $J\to J_c$, $\Delta S \to 0$ and $\cot\theta \to \infty$, becoming nearly vertical. In addition, the red dashed curve schematically indicates what would be expected if the first order transition is slightly broadened by disorder. Note that this is very hard to distinguish from the entropy in a second-order transition. }
    \label{fig:SCv}
\end{figure}

The rest of the paper is structured as follows. We demonstrate the onset of this direct transition from non-magnetic normal state to a $\mathcal{T}$-broken superconductor over an extended parameter range when the coupling $J$ exceeds a threshold using a Landau free energy in  Section \ref{landau}. 
Being a first-order transition~(Fig.~\ref{fig:FreeEn}), this reconciles the experimental phase diagram with thermodynamics. 
Since the superconducting and magnetic order parameters are close to a continuous transition, the transition is weakly first-order and the associated jumps in order parameters and peak in $C_v/T$ may be small as is demonstrated in Fig. \ref{fig:SCv}.

A key question, then, relates to the strength of the cubic coupling $J$ between pairs and local moments in \eqref{eq:PairKondo}.
We present two complementary microscopic derivations of this pair-Kondo interaction in Section \ref{sec:microscopic}, in terms of a weak-coupling picture of single-particle spin-exchange with an S=1/2 local moment and a strong-coupling picture of pair-spin exchange with an S=1 local moment. 
Conventional wisdom~\cite{mineev1999} suggests that this coupling is proportional to the slope of the density of states near the Fermi surface and is therefore, very weak. 
In the weak coupling limit, we reproduce this tendency in Section \ref{secWeakCoupling}. 
Contrary to this conventional expectation, in Section \ref{secStrongCoupling}, we demonstrate a local strong-coupling mechanism in which this coupling strength is governed by atomic energy scales and is \emph{not} small. 
In essence, we describe the two-particle analog of Kondo screening by pairs rather than electrons, and derive a pair-Kondo interaction by a Schrieffer-Wolff decoupling of two-electron valence fluctuations.
 

\section{Phase diagram and weak first order transitions via pair-Kondo coupling}\label{landau}

The presence of two pair potentials in distinct irreps allows for a direct coupling of pairs to local moments as in \eqref{eq:PairKondo}, without pair-breaking. 
Scattering off slowly-fluctuating local moments results in a \emph{gain} in pairing energy for chiral pairs $b_{\Gamma^{\phantom{'}}}\pm i b_{\Gamma^{'}}$, relative to the pairs $b_{\Gamma(\Gamma^{'})}^{\dagger}$ which are eigenstates of $\mathcal{T}$ and point group symmetries.
This allows non-degenerate order parameters $\langle b_{\Gamma(\Gamma^{'})}^{\dagger}\rangle$ to condense simultaneously at a weak first-order transition.

We demonstrate this mechanism by considering the Landau free energy of a system with three order parameters --- two superconducting ones  $\Delta_{1},\Delta_2$  and one corresponding to magnetism, $M$. 
Two of the corresponding transitions are assumed to be close to one another in parameter space. 
Below we focus on the case where one superconducting order $\Delta_{1}$ and a magnetic order $M$ condense at close temperatures $T_{c1},T_{m}$, respectively, while the second superconducting order $\Delta_{2}$ does not condense on its own. 
This assignment models the proximity of UTe$_2$ to the ferromagnetic critical point (in accord with experiments~\cite{ran2019,sundar2019,tokunaga2019,wei2022}) without assuming degeneracy of two superconducting orders. 
We note that our analysis will also apply for a different choice of two critical order parameters. The Landau expansion for free energy takes the form:
\begin{align}
F=&\frac{\alpha_{1}}{2}\left(T-T_{c1}\right)\left|\Delta_{1}\right|^{2}
+\frac{a_2}{2}\left|\Delta_{2}\right|^{2}
+\frac{\alpha_{m}}{2}\left(T-T_{m}\right)M^{2}\notag\\
&-\frac{iJM}{2}\left(\Delta_{1}^{*}\Delta_{2}-\Delta_{1}\Delta_{2}^{*}\right)+F_4\label{eq:FreeEn}
\end{align}
where the cubic coupling in the free energy represents a coupling of the total moment of the Cooper pairs to the proximate magnetization and has been introduced on symmetry grounds in \cite{mineev1999,hillier2012b,nevidomskyy2020} and  
\begin{align}
    F_4=&\frac{b_M}{4} M^4 +\frac{b_1}{4}  |\Delta_1|^4 + \frac{b_2}{4}  |\Delta_2|^4\notag\\
        &+\frac{b_{1 M}}{2} M^2|\Delta_1|^2 
        +\frac{b_{2M}}{2} M^2|\Delta_2|^2
        +\frac{b'_{12}}{2}|\Delta_1|^2|\Delta_2|^2 \notag\\
		&+\frac{b_{12}''}{4} (\Delta_1^2 \Delta_2^{*2}+\Delta_1^{*2}\Delta_2^2)    
\end{align}
where all coefficients are positive so that $F_4 \ge 0$.
With this phenomenological free energy we can study the phase diagram as a function of $T_{c1}$ and $T_m$.

We begin with considering the behavior when $T_{c1}=T_{m}$. 
The individual critical temperatures depend on material parameters such as pressure, strain, and fields and may generically have an accidental crossing. 
At this temperature, which is the bicritical point in absence of the cubic coupling $J=0$, the free energy takes the form 
\begin{equation}
F_{bicr}=\frac{a_2}{2}\left|\Delta_{2}\right|^{2}-\frac{iJM}{2}\left(\Delta_{1}^{*}\Delta_{2}-\Delta_{1}\Delta_{2}^{*}\right)
+F_4.
\end{equation}
The cubic term here favors a $\pi/2$ phase between the order parameters $\Delta_1$ and $\Delta_2$, such that we can take ${\rm Im} \Delta_1 = {\rm Re} \Delta_2 = 0$ without loss of generality. 
As $a_2>0,F_4>0$ it is evident that for sufficiently small $J$ this free energy has only a trivial minimum $\Delta_2=\Delta_1=M=0$. 
However, minimizing $F_{bicr}$ with respect to the order parameters at arbitrary $J$ (see Appendix ~\ref{appDegen} for details) we find that for $J>J_c = \sqrt{a_2(b_{1M}+\sqrt{b_1 b_M})}$, a non-trivial global minimum appears.
Importantly, this minimum splits gradually from the trivial one, such that for $J\to J_c$ one has: $\Delta_1,M \propto \sqrt{J-J_c}, \Delta_2\propto J-J_c$ for a generic choice of parameters \footnote{In the fine-tuned case $b_{12}'=b_{12}'',b_{2M}=0$ the scaling of order parameters is $\Delta_1,M \propto (J-J_c)^{1/4}, \Delta_2\propto \sqrt{J-J_c}$ (see Appendix \ref{appDegen})}. (the full expressions are given in Eq. \eqref{eq:OPscaling}).

While the new minimum appears continuously at $J>J_c$, it leads to a first-order transition as a function of temperature. 
To estimate the jumps in the order parameter at the transition, here we present a general argument, with concrete calculations given in Appendix \ref{appDegen}. 
Close to $J_c$, the free energy at the new minimum scales as $F_{bicr}^{min} \propto -|J-J_c|^3$. 
Increasing the temperature from the erstwhile bi-critical point $T=T_m (=T_{c1})$  introduces an additional positive quadratic term $\propto(T-T_m)  M^2, (T-T_m) \Delta_1^2\propto (T-T_m) (J-J_c)$. 
This implies that the critical temperature at which the new minimum's energy will equal the normal state free energy scales as $T_c - T_m \propto (J-J_c)^2$. 
Such a temperature difference, however, introduces only weak corrections to the order parameters, implying a finite jump to zero at $T_c$.
Indeed $(T-T_m)  M^2 \propto (J-J_c)^3 \ll M^4\propto (J-J_c)^2$; thus, the effect of temperature on the value of the order parameter can be included in a perturbative renormalization of $b_{M},b_1$. 

Thus, there is a finite threshold for the pair-Kondo coupling $J_{c}$ above which the two continuous phase transitions at $T_{c1},T_{m}$ in the vicinity of the accidental crossing, are pre-empted by a single weakly first-order transition, at which both superconducting order parameters develop simultaneously with a discontinuous jump. From the above quoted scaling properties of the order parameters, the jumps in the order parameter at $T_c$ scale as $\delta\Delta_2\propto (J-J_c),\delta(\Delta_1,M)\propto \sqrt{J-J_c}$ as $J \to J_c+0$  (see also Appendix~\ref{appDegen} ). 


We are now in a position to discuss the phase diagram of the model \eqref{eq:FreeEn} in the generic case when $T_{c1}\neq T_m$.
In particular, we consider a linear variation of the underlying second-order transition temperatures indicated by dashed lines in Fig~\ref{fig:PhaseDiag} with a tuning parameter $r$ that stands in for pressure, strain or other perturbation. 
We then find the global minima of the free energy in \eqref{eq:FreeEn} numerically across the phase diagram in Fig \ref{fig:PhaseDiag}.
The red, green and blue levels of the markers are mapped to the normalized values of the three order parameters, so that the disordered phase at high temperature is white and the dark phase at low temperature has all three order parameters finite. 

For an extended range of $r$ in Fig \ref{fig:PhaseDiag}, there is a weakly-first-order transition into the $\mathcal{T}$-broken superconductor (SC2) with a jump in all order parameters. 
As the underlying phase boundaries separate, beyond a critical $r_{c}$, there is a two-step transition - first into a $\mathcal{T}$-invariant single-component superconductor (SC1) and then via a weakly first-order transition where the second component and the concomitant magnetism emerge discontinuously. 
The order parameters' jump across the first order transition decreases as the underlying phase boundaries
separate further, and eventually at large $r$, the first order transition becomes second-order after a critical point (not shown). 

\subsection{Experimental Prediction I: Latent heat}
Above we have demonstrated that the pair-Kondo coupling provides a mechanism capturing the emergence of $\mathcal{T}$-breaking superconductivity from non-degenerate order parameters via a weak first order transition. 
We will now discuss the experimental signatures of such a scenario.

We first discuss the calorimetric signatures of the weak first order transition. 
The latent heat of the first-order transition is given by the discontinuity in the entropy across the transition
$\Delta Q=T_{c}\left(S(T_{c}^{+})-S(T_{c}^{-})\right)$ that corresponds
to integrated weight in the spike in the specific heat $C_{v}/T$
across the transition $\Delta Q/T_{c}=\int_{T_{c}^{-}}^{T_{c}^{+}}dT\frac{C_{v}}{T}$.
We estimate the latent heat to scale as $\left(J-J_{c}\right)^{2}$ and the jump in the specific heat $\Delta\frac{C_{V}}{T}$ to scale as $1/(J-J_c)$ as $J\to J_{c}^+$ (see Appendix \ref{appAnalytical}).
In Fig \ref{fig:SCv} we show the entropy and specific heat across the transition, which is qualitatively similar to the observed specific heat in Ref.~\cite{aoki2022a}.

In such a weakly first-order transition, the small spike in the specific heat $\Delta S$ coexists with a much larger jump in the specific heat $\Delta\frac{C_{V}}{T}$, and may be unobservable, as we emphasize in Fig \ref{fig:SCv}. 
The key prediction of the theory, however, is that as $J\to J_c$ the product approaches a constant. For a simplified case where $\alpha_1(T-T_{c1})=\alpha_M(T-T_{m})\equiv a,\;b_1=b_2=b_M\equiv b,\;b_{12}=b_{1M}=b_{2M}=b'$ (see Appendix \ref{appAnalytical}) we get
\begin{align}
    \Delta S\cdot \Delta\frac{C_{V}}{T} \propto \frac{a_2^2}{b'J_c^2} = \frac{a_2}{b'(b+b')}.
\end{align}

Note that while latent heat is suppressed, the jump in the specific heat diverges $\alpha^{-2} \Delta C_v/T_c \to 2 a_2/(J_c (J - J_c))$ as $J\to J_c$. This will manifest itself as a strong deviation from BCS expectation of $\Delta C_v/T_c = 1.43 \gamma_N$ where $\gamma_N$ is the normal state Sommerfeld coefficient. Furthermore, in the part of the phase diagram where first-order transition occurs at a lower temperature than a second order one, the specific heat jump will be higher at the first-order transition. This contradicts the usual intuition that a transition at lower temperature should be accompanied by a smaller specific heat jump due to the depleted density of states.This is consistent with the observations in UTe$_2$ under pressure \cite{braithwaite2019,aoki2020multiple}. Alternative mechanisms for such behavior involve a state with more nodes on the Fermi surface condensing first \cite{andersen2022heat}. 

\subsection{Experimental Prediction II: Ultrasound signatures}

The change in the propagation of sound through the sample as it undergoes a phase transition places strong constraints on the symmetries of the order parameter. 
In particular, when a two-component superconducting order parameter emerges at a single second order phase transition, there is a characteristic jump in the velocity of a shear mode, whose size is related to the specific heat jump~\cite{sigrist2002}. 
Understanding how the sound velocities evolve across the weakly-first order transition allows us to distinguish its signatures from other candidate scenarios.

The propagation of sound through the sample is governed by the following Lagrangian density:
\begin{equation}
    \mathcal{L}_{el} = \int d^3 {\bf r} \left(\rho \frac{\dot{\bf u}^2}{2} - U_{el}[\{\varepsilon_{\alpha\beta}\}]\right)
\end{equation}
where $\rho$ is the mass density, ${\bf u}({\bf r},t)$ is the displacement of the ion, $\dot{\bf u}=\partial {\bf u}/\partial t$ and the elastic energy density $U_{el}$ is a function of the local strain $\varepsilon_{ij}=\frac{1}{2} \left(\frac{\partial u_i}{\partial r_j}+\frac{\partial u_j}{\partial r_i}\right)$
\begin{align*}
    U_{el}&=
    \left(c_{11}\frac{\varepsilon_{xx}^2}{2}
    +
    c_{22}\frac{\varepsilon_{yy}^2}{2}
    +
    c_{33}\frac{\varepsilon_{zz}^2}{2}
    \right)\\
    &+
    \left(c_{12}\varepsilon_{xx}\varepsilon_{yy}
    +
    c_{23}\varepsilon_{yy}\varepsilon_{zz}
    +
    c_{31}\varepsilon_{zz}\varepsilon_{xx}
    \right)
    \\
    &+c_{44} \frac{\varepsilon_{yz}^2}{2}
    +c_{55} \frac{\varepsilon_{xz}^2}{2}
    +c_{66} \frac{\varepsilon_{xy}^2}{2}
\end{align*}
where $c_{ij}$ are the various elastic constants of an orthorhombic lattice.

Although the theory so far has been completely general with respect to the order parameter symmetry and the nature of the proximate magnetism, to demonstrate the coupling of the order parameters in \eqref{eq:FreeEn} with the local strain, we must specify the irreps under which they transform. 
Since the near-critical magnetism in UTe$_2$ has its easy axis along $x$~\cite{ran2019,aoki2019a}, we assume the magnetic order parameter to be $M_x$ which transforms under the $B_{3g}$ irrep of $D_{2h}$ and the superconducting order parameters $\Delta_{1}$ and $\Delta_{2}$ to transform under $B_{1u}$ and $B_{2u}$ respectively. 
The order parameters in \eqref{eq:FreeEn} then have a linear coupling to strain given by
\begin{align}
&U^{el-OP}_1 = \lambda \varepsilon_{yz} (\Delta_{B_{1u}}^* \Delta_{B_{2u}} + \Delta_{B_{2u}}^* \Delta_{B_{1u}} ) \notag\\
&+
    \sum_\alpha \frac{\varepsilon_{\alpha\alpha}}{2} 
    (\lambda_{\alpha,1}^{(1)}|\Delta_{B_{1u}}|^2
    +\lambda_{\alpha,2}^{(1)}|\Delta_{B_{2u}}|^2 
    +\lambda_{\alpha,M}^{(1)} M_x^2)   \label{eq:lin} 
\end{align}
and a quadratic coupling to strain given by
\begin{align}
&U^{el-OP}_2 =
    \lambda^{(2)} \varepsilon_{xy}\varepsilon_{xz} (\Delta_{B_{1u}}^* \Delta_{B_{2u}} + \Delta_{B_{2u}}^* \Delta_{B_{1u}} )
    \notag\\
    &+
    \sum_\alpha \frac{\varepsilon_{\alpha\alpha}^2}{2} (\lambda_{\alpha,1}^{(2)} |\Delta_{B_{1u}}|^2+\lambda_{\alpha,2}^{(2)} |\Delta_{B_{2u}}|^2+\lambda_{\alpha,M}^{(2)} M_x^2)
    \notag\\
    &+
    \sum_\alpha \varepsilon_{\alpha\alpha}\varepsilon_{\beta\beta} 
    (\lambda_{\alpha\alpha,\beta\beta,1}^{(2)}|\Delta_{B_{1u}}|^2 \notag\\
    &\phantom{+\sum_\alpha \varepsilon_{\alpha\alpha}\varepsilon_{\beta\beta}}
    +\lambda_{\alpha\alpha,\beta\beta,2}^{(2)}|\Delta_{B_{2u}}|^2
    +\lambda_{\alpha\alpha,\beta\beta,M}^{(2)} M_x^2)
    \notag\\
       &+
    \sum_{\alpha\neq\beta} \frac{\varepsilon_{\alpha\beta}^2}{2} (\lambda_{\alpha\beta,1}^{(2)} |\Delta_{B_{1u}}|^2 \notag\\
    &\phantom{\sum_{\alpha\neq\beta} \frac{\varepsilon_{\alpha\beta}^2}{2}}
    +\lambda_{\alpha\beta,2}^{(2)} |\Delta_{B_{2u}}|^2
    +\lambda_{\alpha\beta,M}^{(2)} M_x^2)   
    \label{eq:quadr}
\end{align}
We work in the low-frequency limit $\omega \tau \ll 1$, where $\omega$ is the sound frequency and $\tau$ is the timescale over which the order parameters equilibrate with the lattice. We can then describe the system by the free energy $F=F_{OP}+V (U_{el}+ U_1^{el-OP}+U_2^{el-OP})$ with 
\begin{align}
    F_{OP}=&\frac{a_1}{2} |\Delta_{B_{1u}}|^2 
    + \frac{a_2}{2} |\Delta_{B_{2u}}|^2 
    + \frac{a_m}{2} M_x^2 \notag \\
    &+ J M_x \Im(\Delta_{B_{1u}} \Delta_{B_{2u}}^*) 
    + \frac{b}{4} (|\Delta_{B_{1u}}|^4 + |\Delta_{B_{2u}}|^4 +M_x^4) \notag \\
    &+ \frac{b'}{2} (|\Delta_{B_{1u}}|^2|\Delta_{B_{2u}}|^2 +|\Delta_{B_{1u}}|^2|M_x^2 + |\Delta_{B_{2u}}|^2|M_x^2)  \notag \\
    &+ \frac{b''}{2} ((\Delta_{B_{1u}}^*)^2 \Delta_{B_{2u}}^2 + (\Delta_{B_{2u}}^*)^2 \Delta_{B_{1u}}^2) \label{eqFOP}
\end{align}
where $a_1=\alpha (T-T_{c1}), a_m=\alpha (T-T_m)$ and $V$ is the system volume. 

Minimizing the free energy in the presence of the strain generically leads to a correction to the first-order transition temperature 
\[ T_{c}(\{\varepsilon_{\alpha\beta}\}) = T_{c}^0 +\sum_{\alpha} \gamma_{\alpha} \varepsilon_{\alpha\alpha} +\sum_{\alpha\beta} \gamma'_{\alpha\beta} \varepsilon_{\alpha\beta}^2 + O(\varepsilon^3) \]
where $T_{c}^0$ is the transition temperature at zero strain. The free energy just below the transition temperature can then be expanded as
\begin{align}
    F - F_N =& \Delta S \left(T-T_{c}(\{\varepsilon_{\alpha\beta}\})\right) \notag\\
    &+ \frac{\Delta C_v}{2T_{c}} \left(T-T_{c}(\{\varepsilon_{\alpha\beta}\})\right)^2 + \frac{\Delta c_{44}}{2} \varepsilon_{yz}^2 \ldots \label{eqFmin}
\end{align}
where $F_N$ is the free energy of the normal state.
By taking appropriate derivatives of this free energy, we can identify the jumps in the specific heat, entropy and the $c_{44}$ shear modulus respectively (see Appendix \ref{appRUS} for details)
\begin{align}
    \frac{\Delta C_v}{T_{c}} &\to \alpha^2\frac{ a_2}{J_c (J-J_c)} \notag \\
    \Delta S &\to \alpha\frac{ a_2 (J-J_c)}{2b' J_c}  \notag \\
    \Delta c_{44} &\to \left( -\frac{2 \lambda^2}{b' J_c} + \frac{J_c(\lambda^{(2)}_{\alpha\beta,1}+\lambda^{(2)}_{\alpha\beta,M})}{2b'(b+b')}\right)  (J-J_c)
\end{align}
in the asymptotic limit $J\to J_c$, where $J_c$ is the critical value of the cubic coupling above which the first order transition appears. 
Across this first-order transition, we predict the following discontinuities in the elastic constants measured in resonant ultrasound
\begin{enumerate}
    \item There should be jumps in all nonshear elastic moduli $c_{\alpha\beta}$, $\alpha,\beta=1,2,3$
    \item The magnitude of the jumps should show small ($\propto J-J_c$) deviations from usual Ehrenfest relations $\lambda_{\alpha 1}^{-2} \Delta c_{\alpha\alpha} >  \alpha^{-2} \Delta C_v/T_c$.
    This is due to the small jumps in the order parameters at the weak first-order transition which are quadratically coupled to the strain coefficients in \eqref{eq:quadr}.
    \item In this case, nonlinear coupling to strain can lead to small jumps in all shear moduli because of the terms $
    \sum_{\alpha\neq\beta} \frac{\varepsilon_{\alpha\beta}^2}{2} (\lambda_{\alpha\beta,1}^{(2)} |\Delta_{B_{1u}}|^2
    +\lambda_{\alpha\beta,2}^{(2)} |\Delta_{B_{2u}}|^2
    +\lambda_{\alpha\beta,M}^{(2)} M_x^2)$ in \eqref{eq:quadr}.
    \item These jumps in the shear moduli have the asymptotic Ehrenfest relations as $J\to J_c$
    \begin{align}
        \Delta c_{44} \frac{\alpha a_2}{2b' J_c} &\to
        \frac{\Delta Q}{T_c} \left(\frac{2\lambda^2}{J_c b'} + \frac{J_c(\lambda^{(2)}_{yz,1}+\lambda^{(2)}_{yz,M})}{2b'(b+b')}\right) \notag \\
        \Delta c_{55} \frac{\alpha a_2}{2b' J_c} &\to
        \frac{\Delta Q}{T_c} \left(\frac{J_c(\lambda^{(2)}_{xz,1}+\lambda^{(2)}_{xz,M})}{2b'(b+b')}\right) \notag \\
        \Delta c_{66} \frac{\alpha a_2}{2b' J_c} &\to
        \frac{\Delta Q}{T_c} \left(\frac{J_c(\lambda^{(2)}_{xy,1}+\lambda^{(2)}_{xy,M})}{2b'(b+b')}\right) 
    \end{align}
where $\Delta Q$ is the small ($\propto J-J_c$) latent heat at the weak first order transition. For comparison, in case of an accidental degeneracy between two second order phase transitions at which $\Delta_{B_{1u}}$ and $\Delta_{B_{2u}}$ turn on simultaneously, the corresponding jump discontinuity in the shear modulus would be related to the specific heat jump by the standard Ehrenfest relation~\cite{sigrist2002}
    \begin{align}
        \Delta c_{44}=-\frac{\Delta C_v}{T_c}\left(\frac{\partial T_c}{\partial \epsilon_{yz}}\right)^2 \frac{2b-2b''+b'}{16 b''}
    \end{align}
\end{enumerate}

\subsection{Discussion and an alternate scenario}
A recent preprint~\cite{theuss2023} has reported the results of a pulsed echo ultrasound experiment on ${\rm UTe}_2$, finding no evidence of jumps in any of the shear moduli at the superconducting transition. For a weakly first order transition, as discussed above, the shear moduli can be parametrically small in $J-J_c$, similarly to the latent heat. Therefore, given a limited experimental resolution, a weak first order transition can be consistent with the results of \cite{theuss2023}. On the contrary, for a second-order transition with two accidentally degenerate orders, there is no natural parameter that can make the shear moduli jumps small.
This scenario involving second-order transitions is therefore inconsistent with the observations of \cite{theuss2023}.
We suggest that the simplest way to reconcile the universal observation of two 
transitions under pressure in UTe$_2$ and the observation of Kerr effect 
in split-$T_c$ samples~\footnote{notably reproduction studies of the Kerr effect on these split-$T_c$ samples have not been reported, although a recent paper~\cite{ajeesh2023} has found absence of spontaneous Kerr rotation in other ${\rm UTe_2}$ samples with a single unsplit $T_c$.} is to search for a small latent heat that goes below the experimental resolution as we approach ambient pressure. 
As emphasized in the previous subsection, we do predict a small jump in all shear moduli at the superconducting transition, which becomes, however, vanishingly small for $J$ close to $J_c$. 
We note that the origin of $\mathcal{T}$-broken 
superconductivity in our weakly first-order scenario does involve
superconducting order parameters in two irreps of the non-magnetic point group.

The phenomenological description above reconciles the lack of an
observed second transition in the low pressure data for ${\rm UTe_{2}}$
(Fig.~\ref{fig:PhaseDiag}b). In particular, it is sufficient to have an accidental degeneracy either between one superconducting order and magnetism, or two superconducting orders to create a finite region of the phase diagram where unsplit transition into a $\mathcal{T}$-broken superconductor exists.
In place of magnetism, any symmetry breaking local order that is nearly degenerate with superconducting order can be used to motivate a similar phase diagram with a single weakly-first-order transition.
The experimental data is also consistent with three
near critical order parameters described by the free energy 
\begin{align}
    &F=\frac{\alpha_{1}(T-T_{c1})}{2}\left|\Delta_{1}\right|^{2}
    +\frac{\alpha_2(T-T_{c2})}{2}\left|\Delta_{2}\right|^{2}\notag\\
    &+\frac{\alpha_{m}(T-T_{m})}{2}M^{2}-\frac{iJM}{2}\left(\Delta_{1}^{*}\Delta_{2}-\Delta_{1}\Delta_{2}^{*}\right)
    +F_4 \label{eq:FreeEnalt}
\end{align}
with a crossing between say, $T_{c1}$
and $T_{m}$ as shown in Fig.~\ref{fig:PhaseDiag}(c), if $T_{c2}$ is close enough
that $J^{2}>\alpha_{2}\left(T_{c1}-T_{c2}\right)b$. In this case, the critical
pair-Kondo coupling may be parametrically small if $T_{c1}-T_{c2}\rightarrow0$.
Since the existence of a weakly first order transition depends on
a delicate balance between the Landau free energy coefficients and
the pair Kondo coupling, it is important to understand the microscopic
origins of the symmetry allowed coupling constant $J$ introduced in \eqref{eq:PairKondo}.

\section{Microscopic origin of pair-kondo coupling}\label{sec:microscopic}

We discuss the microscopic mechanism of pair-Kondo coupling, that we introduced phenomenologically in the preceding Section. 
In particular, we identify a mechanism by which Cooper pairs can change their symmetry representations by scattering off local moments. 
Here we consider two cases. In the weak pairing or BCS limit, where $T_c\sim \Delta_{1,2}$, the change of symmetry proceeds via breaking of one pair, spin-scattering that rotates one of the two electron-spins, and recombination into another pair. Alternatively, in the strong pairing limit,
via coupling to an local moment, the pair spin itself can rotate in the
scattering process, as we demonstrate for an S=1 moment. In presence of spin-orbit coupling and crystal
fields, the intuition from spin-rotation translates to a change in
the irreducible representation of the Cooper pair. We capture this
intuition by explicitly calculating the pair-Kondo coupling from microscopic
Hamiltonians in the following subsections.

\subsection{Weak coupling to S=1/2 moment}\label{secWeakCoupling}

In a weak-pairing superconductor, the energy gain by pairs coupling to moments should be proportional to $\mathbf{S}\cdot \mathbf{m}_{\rm SC}$ where $\mathbf{S}$ is the total spin of the moments, and $\mathbf{m}_{\rm SC}$ is the magnetic moment of the superconductor in absence of local moments. 
The latter is known~\cite{fomin1978a,leggett1975,mineev1999} to be proportional to the slope of the normal state density of states - if the density of states is constant over the superconducting gap scale, the $\mathcal{T}$-breaking non-unitary superconductor has no net moment. 
Here we show that this intuition also holds for order parameters above the second-order critical temperatures - in the weak pairing limit, the coefficient of the cubic coupling between two superconducting orders and magnetism scales with the slope of the density of states. 
Starting from a Hamiltonian with generic interactions that drive pairing in two channels and magnetism in a third channel, the goal is to first decouple the interactions by introducing auxiliary fields for each ordering tendency and then examine the lowest order coupling between the three fields in the free energy.

Consider the model Hamiltonian
\[ H=H_K + H_1 + H_2 + H_M \]
where $H_K=\sum_{\bfk,\alpha} \xi_\bfk c_{\bfk a}^{\dagger} c_{\bfk \alpha}^\phdag$ is the kinetic energy of non-interacting fermions on an arbitrary lattice with crystal momentum $\bfk$ and spin $\alpha$ created by $c_{\bfk\alpha}^\dagger$,
\[H_{1}=-\sum_{\bfk,\bfk'} |V_1| c_\bfk^{\dagger} \hat{\phi}_{1 \bfk} c_{-\bfk}^{\dagger} c^\phdag_{-\bfk^{\prime}} \hat{\phi}_{1 \bfk^{\prime}} c^\phdag_{\bfk^{\prime}} 
\equiv - |V_1| b_{1}^\dagger b_{1}\]
is an attractive interaction for fermion pairs created by $b_{1}^\dagger \equiv \sum_{\bfk\alpha\beta} c_{\bfk\alpha}^{\dagger} \hat{\phi}_{1 \bfk,\alpha\beta} c_{-\bfk\beta}^{\dagger}$ with a hermitian form-factor $\hat{\phi}_1$ that is a matrix in spin space ($\alpha,\beta = \{\uparrow,\downarrow\}$ ) and transforms according to a one-dimensional irrep of the point group,
\[H_{2}=- \sum_{\bfk,\bfk'} |V_2|
c_\bfk^{\dagger} \hat{\phi}_{2 \bfk} c_{-\bfk}^{\dagger} 
c_{-\bfk'} \hat{\phi}_{2 \bfk'} c_{\bfk'}
\equiv - |V_2| b_{2}^\dagger b_{2}\]
is an attractive pairing interaction in an orthogonal channel transforming under a different one-dimensional irrep,
\[H_{M}=\sum_{\bfk,\bfk',\bfq} V_M(\bfq) 
c_{\bfk+\bfq}^{\dagger} \boldsymbol{\sigma} c_\bfk^\phdag
\cdot 
c^\dagger_{\bfk^{\prime}-\bfq}  \boldsymbol{\sigma} c_{\bfk^{\prime}}^\phdag
\equiv \sum_\bfq V_M(\bfq) S_{\bfq} \cdot S_{-\bfq}\]
is a generic local magnetic exchange interaction that favors ferromagnetism: $V_M(\bfq=0)<0$. 
Since we are interested in systems close to a ferromagnetic instability, we ignore finite momentum spin scattering and set $V_M(\bfq\neq0)=0$.

Decoupling the interactions using Hubbard-Stratonovich fields $\Delta_1,\Delta_2,\bf{M}$ allows us to express the partition function $\mathcal{Z}=\int_{\Delta_1,\Delta_2,\bf{M},\psi} e^{-S}$ in terms of an action that is quadratic in the fermionic fields
\[S = -T\sum_{\bfk,n} 
\Psi_{\bfk,n}^{\dagger}
G_{\bfk,n}^{-1}
\Psi_{\bfk,n} + O(|\Delta_1|^2,|\Delta_2|^2,M^2)\]
where the Nambu spinor $\Psi_{{\bf k}}^{\dagger}=\left(\begin{array}{cc}
\bar{\psi}_{{\bf k}} & \psi_{-{\bf k}}\Theta\end{array}\right)$ contains the Grassmann fields of the fermions, $\Theta=i\sigma_y \mathcal{K}$ (with $\mathcal{K}$: complex conjugation) is the time-reversal operator, and the inverse of the Nambu-Gorkov Green's function
$G_{{\bf k},n}^{-1}=G_{0{\bf k},n}^{-1}-\Sigma_{{\bf k}}$ can be
decomposed into a bare inverse Green's function $G_{0{\bf k}}^{-1}=i\omega_{n}-\xi_{{\bf k}}\tau_{3}$
and the self energy 
\begin{equation}
\Sigma_{{\bf k}}=\left(\begin{array}{cc}
{\bf M}\cdot\boldsymbol{\sigma} & \Delta_{1}\hat{\phi}_{1{\bf k}}+\Delta_{2}\hat{\phi}_{2{\bf k}}\\
\Delta_{1}^{*}\hat{\phi}_{1{\bf k}}+\Delta_{2}^{*}\hat{\phi}_{2{\bf k}} & {\bf M}\cdot\boldsymbol{\sigma}
\end{array}\right)
\end{equation}
where $\omega_{n}=(2n+1)\beta/2$ are fermionic Matsubara frequencies and $\tau_{i}$ are Pauli matrices in Nambu space. 
Integrating out the fermionic fields, we arrive at the free energy at the saddle point
\begin{equation}
F=T\sum_{{\bf k},n}\Tr\log G_{{\bf k},n}^{-1}=F_{N}+T\sum_{{\bf k},n}\Tr\log\left(1+G_{0{\bf k},n}\Sigma_{{\bf k}}\right)
\end{equation}
where $F_{N}=T\sum_{{\bf k},n}\Tr\log G_{0{\bf k},n}^{-1}$ is the normal state free energy. 
The leading order terms in the Taylor expansion
of the logarithm that survives the Nambu trace and the sum over the
Brillouin zone are $F=F_{0}+F_{2}+F_{3}+...$ 
\begin{widetext}
\begin{equation*}
F_{2}=T\sum_{{\bf k},n}\Tr\left[\frac{|\Delta_{1}|^{2}\hat{\phi}_{1{\bf k}}^{2}+|\Delta_{2}|^{2}\hat{\phi}_{2{\bf k}}^{2}}{\left(i\omega_{n}-\epsilon_{{\bf k}}\right)\left(i\omega_{n}+\epsilon_{{\bf k}}\right)}
+M^{2}\left(\frac{1}{\left(i\omega_{n}-\epsilon_{{\bf k}}\right)^{2}}+\frac{1}{\left(i\omega_{n}+\epsilon_{{\bf k}}\right)^{2}}\right)\right]
\end{equation*}

\begin{align*}
F_{3} & =-T M_{m}\sum_{{\bf k},n}\Tr\sigma_{m}\frac{\Delta_{1}\Delta_{2}^{*}\hat{\phi}_{1{\bf k}}\hat{\phi}_{2{\bf k}}+\Delta_{1}^{*}\Delta_{2}\hat{\phi}_{2{\bf k}}\hat{\phi}_{1{\bf k}}}{\left(i\omega_{n}-\epsilon_{{\bf k}}\right)^{2}\left(i\omega_{n}+\epsilon_{{\bf k}}\right)}+\frac{\Delta_{1}\Delta_{2}^{*}\hat{\phi}_{2{\bf k}}\hat{\phi}_{1{\bf k}}+\Delta_{1}^{*}\Delta_{2}\hat{\phi}_{1{\bf k}}\hat{\phi}_{2{\bf k}}}{\left(i\omega_{n}-\epsilon_{{\bf k}}\right)\left(i\omega_{n}+\epsilon_{{\bf k}}\right)^{2}}\nonumber \\
 & =-T M_{m}\sum_{{\bf k},n}\Tr\sigma_{m}\frac{\Im\left[\Delta_{1}\Delta_{2}^{*}\right]\left(\hat{\phi}_{1{\bf k}}\hat{\phi}_{2{\bf k}}-\hat{\phi}_{2{\bf k}}\hat{\phi}_{1{\bf k}}\right)}{\left(i\omega_{n}-\epsilon_{{\bf k}}\right)\left(i\omega_{n}+\epsilon_{{\bf k}}\right)}\left(\frac{1}{\left(i\omega_{n}-\epsilon_{{\bf k}}\right)}-\frac{1}{\left(i\omega_{n}+\epsilon_{{\bf k}}\right)}\right)
\end{align*}
where $M_{m}={\bf M}_{\Gamma_{1}\times\Gamma_{2}},$ the component of the three-vector
that transforms under the product irrep $\Gamma_{1}\times\Gamma_{2}$ of the pairing order parameters.
The hermitian form-factors $\hat{\phi}_{1,2}$ can be expanded in the basis
of Pauli matrices with real coefficients $\hat{\phi}_{1{\bf k}}=\phi_{1{\bf k}}^{0}\sigma_{0}+{\bf d}_{1{\bf k}}\cdot\boldsymbol{\sigma}$,
so that the lowest order coupling between the magnetic order parameter and the two pairing order parameters takes the form
\begin{equation}
F_{3}=-J_{PK}M_{m}\Im\left[\Delta_{1}\Delta_{2}^{*}\right]
\end{equation}
with \[J_{PK}=T\sum_{{\bf k},n}\left({\bf d}_{1{\bf k}}\times{\bf d}_{2{\bf k}}\right)\cdot \hat{m}\sum_{n}\left[\left(i\omega_{n}-\epsilon_{{\bf k}}\right)^{-2}\left(i\omega_{n}+\epsilon_{{\bf k}}\right)^{-1}-\left(i\omega_{n}-\epsilon_{{\bf k}}\right)^{-1}\left(i\omega_{n}+\epsilon_{{\bf k}}\right)^{-2}\right].\]
Evaluating the Matsubara summations gives a closed form expression
for the coupling constant 
\begin{equation}
J_{PK}=\sum_{{\bf k}}\hat{m}\cdot\left({\bf d}_{1{\bf k}}\times{\bf d}_{2{\bf k}}\right)\left[\frac{2}{\epsilon_{{\bf k}}}\left(\frac{df}{d\epsilon}|_{\epsilon_{{\bf k}}}+\frac{df}{d\epsilon}|_{-\epsilon_{{\bf k}}}\right)-\frac{2}{\epsilon_{{\bf k}}^{2}}\left(f(\epsilon_{{\bf k}})-f(-\epsilon_{{\bf k}})\right)\right].
\end{equation}
To simplify this expression, we assume that the angular average of the d-vectors $D_{FS}(\epsilon)=\int_{FS(\epsilon)} d\Omega \left({\bf d}_{1{\bf k}}\times{\bf d}_{2{\bf k}}\right)$ is smooth near the Fermi surface, and evaluate
\begin{align}
    J_{PK}&=2 D_{FS}^0 \int_{-D}^D d\epsilon N(\epsilon) \left[\frac{2}{\epsilon_{{\bf k}}}\left(\frac{df}{d\epsilon}|_{\epsilon_{{\bf k}}}+\frac{df}{d\epsilon}|_{-\epsilon_{{\bf k}}}\right)-\frac{2}{\epsilon_{{\bf k}}^{2}}\left(f(\epsilon_{{\bf k}})-f(-\epsilon_{{\bf k}})\right)\right] \label{eqJPK}
\end{align}
where $D$ is the bandwidth, $N(\epsilon)$ is the density of states and $D_{FS}^0=D_{FS}(E_F)$ is the Fermi surface average of the pair moment. The first term can be integrated by parts to give
\begin{align}
    \int d\epsilon N(\epsilon) \frac{f'(\epsilon)}{\epsilon} = N(\epsilon) \frac{f(\epsilon}{\epsilon} \bigg|_{-D}^D - \int d\epsilon  f(\epsilon) 
    \left[ \frac{\partial N(\epsilon)}{\epsilon \partial \epsilon} - \frac{N(\epsilon)}{\epsilon^2} \right]
\end{align}
Using this in \eqref{eqJPK}, we find 
\begin{align*}
    J_{PK}= 2 D_{FS}^0  \left( \frac{N(\epsilon) \tanh \beta \epsilon/2}{\epsilon}\bigg|^{D}_{-D} - \int_{-D}^D  \frac{d\epsilon}{\epsilon} \left( f(\epsilon)N'(\epsilon) + f(-\epsilon)N'(-\epsilon) \right) \right)
\end{align*}
For a wide band where the first term vanishes, the cubic coupling $J_{PK}$ is proportional to the slope of the density of states near the the Fermi level, at any saddle point where three order-parameters are finite, even for local minima of the free energy landscape. This is consistent with the vanishing magnetic moment of a non-unitary superconductor (c.f. pg 105 in \cite{mineev1999}), corresponding to the case where the saddle point \emph{is} the global minimum. 
\end{widetext}

\subsection{Strong Coupling to S=1 moment}\label{secStrongCoupling}

Pairs can also couple to local moments by mediating charge-2e valence
fluctuations, similar to how electrons Kondo-couple to moments by
mediating charge-e valence fluctuations. We demonstrate this process
for a spin-1 $f^{2}$ local moment with three states $|S_{x}=0\rangle,|S_{y}=0\rangle,|S_{z}=0\rangle$
in the SU(2)-symmetric ground state manifold. 

The kinetic energy of the conduction electrons is described by the generic Hamiltonian
\begin{align}
    H_K=\sum_{\bfk\alpha} \epsilon_\bfk^c c_{\bfk \alpha}^\dagger c_{\bfk \alpha} + \sum_{\bfq,m} \epsilon_{\bfq,m}^b b_{\bfq m}^\dagger b_{\bfq m} 
\end{align}
with $b_{\bfq m}=N^{-1/2}\sum_{j} b_{jm}^{\dagger}e^{i \bfq \cdot {\bf r}_j}$, and $b_{jm}^{\dagger}=\sum_{\boldsymbol{\delta}\alpha\beta}c_{j\alpha}^{\dagger}\hat{\phi}_{m\boldsymbol{\delta},\alpha\beta}c_{j+\boldsymbol{\delta},\beta}^{\dagger}$, where $j$ runs over unit cells, $\alpha,\beta$ are indices for internal electronic degrees of freedom, including orbital and spin and $\boldsymbol{\delta}$ is a lattice displacement vector. $b_{jm}^{\dagger}$
creates a pair with the same local symmetry as a triplet pair with its d-vector along $m$. 
$c_{\bfk\alpha}^{\dagger}$ creates a conduction electron in a Bloch wave of orbital $\alpha$ (including spin) with momentum $\bfk$ and dispersion $\epsilon^c_\bfk$. 
The pairs have the dispersion $\epsilon_{\bfq}^b$ which is negative at $\bfq=0$. 
The coupling of pairs to local moments that we derive below is independent from the origin of the attraction between pairs leading to $\epsilon_{\bfq=0}^b<0$.
For concreteness, we work in the strong coupling regime, where the single particle spectrum is fully gapped, even in the absence of superfluid stiffness when the pairs have no off-diagonal long-range order, i.e. above the superconducting $T_c$. The only valence fluctuations of the local moment are then mediated by the pairs at low energy.

Charge fluctuations of the local moment are captured by the mixed-valence Hamiltonian 
\begin{equation}
H=\sum_{j}\Bigg(E_{0}X^{00}(j)+\sum_{m=\{x,y,z\}}\left(VX_{m}^{02}(j)b_{jm}^{\dagger}+{\rm h.c.}\right)\Bigg)\label{eq:2eValence}
\end{equation}
where the Hubbard operator $X_{m}^{02}=|f^{0}\rangle\langle f_{m}^{2}|$
removes two f electrons from the $S_{m}=0$ state of the ground state
triplet.

In the limit $V/E_{0}\ll1$, the perturbative effect of virtual valence
fluctuations on the ground state manifold is captured by a Schrieffer-Wolff
transformation after which the Hamiltonian $e^{iS}He^{-iS}$ is block-diagonal
to second order in $V/E_{0}$. This is effected by $iS=\sum_{jm}E_0^{-1}\left(V X_{m}^{02}(j)b_{jm}^{\dagger}-{\rm h.c.}\right)$.
The transformed Hamiltonian is then 
\begin{align}
e^{iS}He^{-iS}=&\sum_{j}
\frac{\left|V\right|^{2}}{E_{0}}\left[X^{00}(j)b_{jm}^{\dagger}b_{jm} 
-X_{mm'}^{22}(j)b_{jm}b_{jm'}^{\dagger}\right]\notag\\
&+\sum_{j} E_{0}X^{00}(j)
+O\left(\left(V/E_{0}\right)^{3}\right)\label{eq:SWHam}
\end{align}
The second term can be simplified using the Fierz identity for Gell-Mann
matrices~\cite{cheng1994} $\delta_{\alpha\beta}\delta_{\gamma\delta}=\frac{1}{3}\delta_{\alpha\delta}\delta_{\gamma\beta}+\frac{1}{2}\boldsymbol{\lambda}_{\alpha\delta}\cdot\boldsymbol{\lambda}_{\gamma\beta}$. Thus 
\begin{align}
X_{mm'}^{22}b_{m}b_{m'}^{\dagger} & =X_{mm'}^{22}b_{n}b_{n'}^{\dagger}\delta_{mn}\delta_{m'n'}\nonumber \\
 & =\frac{1}{3}b_{m}b_{m}^{\dagger}+\frac{1}{2}|f_{m}^{2}\rangle\boldsymbol{\lambda}_{mm'}\langle f_{n}^{2}|\cdot b_{n}\boldsymbol{\lambda}_{nn'}b_{n'}^{\dagger}
\end{align}
where we can identify the three imaginary Gell-Mann matrices as the
S=1 generators of SU(2) following Ref.~\cite{wang1997} $S_{l}=i|f_{m}^{2}\rangle\epsilon_{lmn}\langle f_{n}^{2}|$.
Projecting the canonically transformed Hamiltonian \eqref{eq:SWHam}
to the low energy spin-sector of the f-electrons, and ignoring potential
scattering of pairs, then gives the pair-Kondo Hamiltonian 
\begin{equation}
H_{PK}=J_{PK}\sum_{jlmn}i\epsilon_{lmn}S_{l}(j)b_{jm}^{\dagger}b_{jn}\label{eq:PairKondolocal}
\end{equation}
with $J_{PK}=\frac{\left|V\right|^{2}}{2E_{0}}.$ If we restrict ourselves to zero-momentum pairing in a translationally invariant system, and recall that $b_{jm}^{\dagger}=\sum_{\boldsymbol{\delta}}c_{j}^{\dagger}\hat{\phi}_{m\boldsymbol{\delta}}c_{j+\boldsymbol{\delta}}^{\dagger}$ then this takes the form 
\begin{equation}
H_{PK}=J_{PK}\sum_{jlmn}i\epsilon_{lmn}{\bf S}_{l,q=0}\sum_{{\bf k,k'}}c_{{\bf k}}^{\dagger}\hat{\phi}_{m{\bf k}}c_{{\bf -k}}^{\dagger}c_{-{\bf k}'}\hat{\phi}_{n{\bf k'}}^{*}c_{{\bf k'}} \label{eq:pairKondokSpace}
\end{equation}
with $\hat{\phi}_{m{\bf k}}=\sum_{\delta}\hat{\phi}_{m\boldsymbol{\delta}}e^{-i{\bf k}\cdot\boldsymbol{{\bf \delta}}}/\sqrt{N}$, describing a change in symmetry of the pairs via scattering off the uniform component of the magnetization.

How does this idealized derivation in the presence of full SU(2) rotational symmetry relate to the experimental reality of UTe$_2$? In general, the presence of spin-orbit coupling in the local moment allows the moment to couple to the crystal fields which break the full rotational invariance. The states $|S_m=0\rangle$ should then be replaced by the crystal field eigenstates that transform under the same irrep as the spin-component $S_m$. In UTe$_2$, the general consensus is that the uranium valence is intermediate between an $f^3$ Kramer's doublet configuration and one or more nearly degenerate $f^2$ singlet configurations, although the relative weight in each valence sector is under debate~~\cite{thomas2020,miao2020,fujimori2021,rosa2022}. For low-symmetry systems like UTe$_2$, two nearly degenerate $f^2$ states is sufficient for an anisotropic coupling to local two-component pairs. The Hubbard operator that connects these two states then transforms like one component of spin, say $S_x=i|f^2_{B_{1u}}\rangle\langle f^2_{B_{2u}}|-i|f^2_{B_{2u}}\rangle\langle f^2_{B_{1u}}|)$, so that \eqref{eq:pairKondokSpace} is modified to describe coupling of one component of magnetism $S_x$ to two pairing components $\Delta_{B_{1u}}$,$\Delta_{B_{2u}}$ precisely as in \eqref{eqFOP}.

Note that as formulated, the pair-Kondo coupling is not restricted only to triplet pairs, but applies quite generally to pairs which have local structure in orbital or sublattice space, captured by the matrix form-factor $\hat{\phi}_{m\boldsymbol{\delta}}$. Thus even in systems where superconductivity is likely to be spin-singlet, 
pairs can couple to local moments by scattering between two or more components with non-commuting form-factors in \eqref{eq:pairKondokSpace} in orbital or sublattice space, providing a mechanism for time-reversal breaking superconductivity beyond ${\rm UTe}_2$.

This mechanism might explain the presence of the frozen moments in early samples~\cite{sundar2023} and the absence of any internal fields in newer high-quality molten-salt-flux grown samples~\cite{azari2023b}. At low concentrations of magnetic impurities, pairs are able to screen the moments via pair-Kondo coupling, but as the concentration of impurity moments increases, long-range interactions between them is expected to result in glassy regimes, similar to the case of magnetic impurities in metals~\cite{mydosh1993}.

\section{Conclusion} 

In this work we presented a mechanism where time-reversal symmetry breaking superconductivity can onset in a single transition in a system without degeneracies between pairing channels. The cubic coupling of Cooper pair magnetic moments to localized magnetic moments results in a weakly first order transition into a spontaneously  $\mathcal{T}$-broken phase preempting individual second order transitions in an extended regime of the phase diagram (Fig. \ref{fig:PhaseDiag}). Remarkably, increasing separation between the bare second-order transition temperatures leads to a splitting of a single transition into two - a second-order one at a higher temperature followed by a weak first order one at low temperature. All of these features are in agreement with the reported behavior of UTe$_2$ under pressure \cite{braithwaite2019,aoki2020multiple}. We have provided further experimental signatures of the proposed scenario in calorimetric and ultrasound measurements. Remarkably, the latent heat for the weak first order transition can be continuously tuned to zero, while the corresponding specific heat jump concomitantly diverges. This offers an alternate explanation for violation of the weak-coupling BCS ratio, while explaining the non-observation of latent heat and shear modulus discontinuities in UTe$_2$. Finally, we discussed the microscopic mechanisms of pair-Kondo coupling in Sec. \ref{sec:microscopic},  in the context of weak and strong coupling, demonstrated for $S=1/2$ moments and $S=1$ local moments respectively. Importantly, while for magnetism of purely itinerant electrons the pair-Kondo coupling is expected to be proportional to the energy derivative of the density of states and thus weak, we find no such suppression for the case of local moments being present.

The mechanism for $\mathcal{T}$-broken superconductivity via pair-Kondo coupling between non-degenerate pairs and local moments is applicable quite generally beyond UTe$_2$, and is not restricted to triplet pairing. 
In a single-band model, singlet pairs in different irreps do not couple to local moments because the pairing form-factors in \eqref{eq:pairKondokSpace} commute. However, in multi-orbital systems, 
the pairing form-factor can have structure in orbital space that allows even spin-singlet pairs to couple to local moments. This allows for a general scenario of weak first order transition into $\mathcal{T}$-broken superconducting phase by the mechanism described above to be applicable to a potentially wider range of materials. Furthermore, our microscopic analysis indicates that systems where itinerant electrons coexist with local moments may be promising for realization of pair-Kondo coupling effects.

Our results can also be straightforwardly extended to the case where instead of magnetic order other particle-hole order is considered, e.g. a cubic coupling between nematic, $s$ and $d$-wave superconducting orders is allowed by symmetry. Other possible combinations allowing for cubic coupling would be uniform superconducting order, charge- or spin- density wave and pair density wave order parameters, that may also have relevance to ${\rm UTe}_2$~\cite{aishwarya2023,gu2023detection,aishwarya2023visualizing}. Since unconventional superconductivity in most cases occurs in systems close to other instabilities~\cite{stewart2017unconventional}, our study provides a new possibility of weak first order transitions to occur in these systems.

\textbf{Acknowledgements:} We are grateful to J\"org Schmalian, Erez Berg, Aline Ramires, Piers Coleman, Priscila Rosa, Mohit Randeria and Daniel Agterberg for very useful discussions on this issue. TH was supported by the U.S. National Science Foundation Grant No. DMR-1830707 and by the Alexander von Humboldt Foundation.

\bibliography{pairKondo}

\appendix

\section{Analytical solution of the Landau free energy}
\label{appDegen}

In this section, we derive the asymptotic expressions for the jumps in the order parameters, the specific heat and the latent heat at the point where the underlying second order transitions at $T_{c1}$ and $T_m$ are degenerate. These analytical expressions hold when the cubic coupling between magnetism and superconducting orders is near the critical coupling $J\to J_c$, and when the biquadratic couplings $b_{12}',b_{12}'',b_{1M},b_{2M}$ in \eqref{eq:FreeEn} are much smaller than the quartic couplings $b_{1,2,M}$.

Starting with the most general form of the Landau free energy describing these three order parameters
\begin{widetext}
\begin{align}
F[\Delta_1,\Delta_2,M] &= \frac{a_1(T,p)}{2} |\Delta_1|^2+ \frac{a_2(T,p)}{2}|\Delta_2|^2 +\frac{a_M(T,p)}{2} M^2  -  \frac{i J}{2} M (\Delta_1^* \Delta_2 - \Delta_1 \Delta_2^*) +\frac{b_M}{4} M^4 +\frac{b_1}{4}  |\Delta_1|^4 + \frac{b_2}{4}  |\Delta_2|^4 \notag\\
&+\frac{b_{1 M}}{2} M^2|\Delta_1|^2 +\frac{b_{2M}}{2} M^2|\Delta_2|^2+\frac{b'_{12}}{2}|\Delta_1|^2|\Delta_2|^2
+\frac{b_{12}''}{4} (\Delta_1^2 \Delta_2^{*2}+\Delta_1^{*2}\Delta_2^2),
\label{eq:GLfull}
\end{align}
where $T$ is the temperature and $p$ is an arbitrary symmetry-allowed tuning parameter, such as pressure. We assume that only two order parameters are close to transition and treat the third one as inert:
\begin{equation}
a_1(T,p)\approx \alpha_1(p) (T-T_1(p)); \,\,\,\,
a_M(T,p)\approx \alpha_M(p) (T-T_M(p)); \,\,\,\,
a_2(T,p) = {\rm const}
\label{eq:GLlinT}
\end{equation}
where one can neglect the variations in $\alpha_{1,M}$ for small variations of the tuning parameter $p$.

Without loss of generality, we can take $\Delta_1$ to be real. The cubic term then in \eqref{eq:GLfull} explicitly favors the imaginary part of $\Delta_2$, resulting in ${\rm Re} \Delta_2=0$ in equilibrium, so that
\begin{align}
F =& \frac{a_1(T,p)}{2} \Delta_1^2+ \frac{a_2(T,p)}{2}\Delta_2^2 +\frac{a_M(T,p)}{2} M^2 
+ J M \Delta_1 \Delta_2 +\frac{b_M}{4} M^4 +\frac{b_1}{4}  \Delta_1^4 + \frac{b_2}{4} \Delta_2^4\notag\\
&+\frac{b_{1 M}}{2} M^2\Delta_1^2 +\frac{b_{2 M}}{2}M^2\Delta_2^2+\frac{b_{12}}{2}\Delta_1^2\Delta_2^2
\label{eq:GLfullRe}
\end{align}
\end{widetext}
where $b_{12} \equiv b_{12}'-b_{12}''$ and we used ${\rm Re} \Delta_1 \to \Delta_1;\;{\rm Im} \Delta_2 \to \Delta_2$ for brevity. Written in this way, the free energy is explicitly invariant if $2$ and $M$ are interchanged in all the coefficients. Therefore, qualitatively similar phase diagrams (in terms of energies and transition temperatures) are expected for the near-degeneracy between SC1 and SC2 or SC1 and M. 


\subsection{Order parameter scaling}

We now set $a_1=a_M=0$ to find the scaling of the order parameters as $J\to J_c$. We start with simplified case where biquadratic couplings $b_{ij}$ are neglected. In this case, the condition the minima in the free energy by solving $\partial F/\partial\Delta_1 = \partial F/\partial\Delta_2 = \partial F/\partial M = 0$ is as follows
\begin{equation}
	\begin{gathered}
b_M M^3+J\Delta_1 \Delta_2 = 0,
\\
b_1 \Delta_1^3+JM \Delta_2 = 0,
\\
a_2 \Delta_2 +b_2 \Delta_2^3 +J M \Delta_1 = 0.
\label{eq:degpoint}
	\end{gathered}
\end{equation}
From the first two equations one finds $\sqrt{b_1 b_M}(\Delta_1 M)=\pm J \Delta_2$ and $\Delta_1/M = (b_M/b_1)^{1/4}$. Substituting in the third one, we get a condition for the existence of a nonzero solution:
\begin{equation}
	J>J_c=\sqrt{a_2\sqrt{b_1b_M}}.
\end{equation}
For $J^2\approx a_2\sqrt{b_1b_M}$ one can also compute the order parameter values:
\begin{equation}
	\begin{gathered}
|\Delta_2| =\frac{\sqrt{J^2-\sqrt{b_1b_M}a_2}}{\sqrt{b_2\sqrt{b_1b_M}}} \propto \sqrt{J-J_c},
\\
|\Delta_1| =  (b_1^3b_M)^{-1/8} \sqrt{J|\Delta_2|}\propto (J-J_c)^{1/4},
\\
|M| = (b_M^3b_1)^{-1/8} \sqrt{J|\Delta_2|} \propto (J-J_c)^{1/4},
	\end{gathered}
 \label{eq:app:nobiq}
\end{equation}
where the signs are such that ${\rm sign}[\Delta_1\Delta_2M]<0$. Most importantly, $\Delta_2\propto \Delta_1 M\ll M,\Delta_1$ close to $J_c$.

The full equations including biquadratic terms takes the form:
\begin{equation}
	\begin{gathered}
		b_M M^3+J\Delta_1 \Delta_2 +  b_{1M}\Delta_1^2 M +b_{2M}\Delta_2^2 M = 0,
		\\
		b_1 \Delta_1^3+J M \Delta_2 + b_{1M}\Delta_1 M^2+ b_{12}\Delta_2^2 \Delta_1  = 0,
		\\
		a_2 \Delta_2 +b_2 \Delta_2^3 +J M \Delta_1 + b_{2M}\Delta_2 M^2+b_{12}\Delta_2 \Delta_1^2 = 0,
	\end{gathered}
\label{eq:degpointfull}
\end{equation}
The first two equations do not allow a simple expression for $\Delta_1 M$ now due to $b_{1M},b_{2M},b_{12}$; however, we can study perturbatively their effects. In lowest order $b_{1M}\ll b_1,b_M$, one can use $\Delta_1/M = (b_M/b_1)^{1/4}$, which results in same equations as \eqref{eq:degpoint}, but with a renormalization $b_1 \to \tilde{b}_1 \approx b_1+b_{1M}\sqrt{b_1/b_M};\;b_M \to \tilde{b}_M \approx b_M+b_{1M}\sqrt{b_M/b_1}$. $b_{12}$ and $b_{2M}$ introduce sub-leading correction, which are however important for what follows next.

The third equation in \eqref{eq:degpointfull}, after substituting the perturbative solution of the first two is modified in a more substantial way:
\begin{align}
    &\left( a_2 - \frac{J^2}{\sqrt{b_1b_M}} \right)\Delta_2 -\frac{3Jb_{2M}}{8} \left(\frac{3}{(b_M^3b_1)^{1/4}} - \frac{1}{\sqrt{b_1b_M}}\right) \Delta_2^2 \notag\\
	&-\frac{3Jb_{12}}{8} \left(\frac{3}{(b_1^3b_M)^{1/4}} - \frac{1}{\sqrt{b_1b_M}}\right) \Delta_2^2= 0,
\end{align}
where $\Delta_2<0$ is assumed, and the $O(\Delta_2^3)$ term is dropped. The leading term in the absence of biquadratic coupling is actually small. While the critical value of coupling remains the same, the scaling of the order parameters at $J\approx J_c$ is not (cf. Eq. \eqref{eq:app:nobiq}):
\begin{equation}
	\begin{gathered}
		|\Delta_2| \propto(J-J_c),
		\\
		|\Delta_1| =  (\tilde{b}_1^3\tilde{b}_M)^{-1/8} \sqrt{J|\Delta_2|}\propto (J-J_c)^{1/2},
		\\
		|M| = (\tilde{b}_M^3\tilde{b}_1)^{-1/8} \sqrt{J|\Delta_2|} \propto (J-J_c)^{1/2}.
	\end{gathered}
\label{eq:deglimitfull}
\end{equation}

Assuming this scaling behavior allows one to calculate the prefactors in \eqref{eq:deglimitfull}. In particular, neglecting the $\Delta_2^3$ term in the third equation in \eqref{eq:degpointfull}, we get:
\begin{align}
\Delta_2 =& -\frac{J M \Delta_1}{a_2+b_{2M} M^2+b_{12}\Delta_1^2} \notag\\
 \approx&  
 -\frac{J M \Delta_1}{a_2}
 +
 \frac{J M \Delta_1(b_{2M} M^2+b_{12}\Delta_1^2)}{a_2^2}.
 \label{eq:del2}    
\end{align}
 Note that to get the correct scaling we need to keep subleading terms of order $(J-J_c)^2$. Using the expansion \eqref{eq:del2} for the first two equations in \eqref{eq:degpointfull} we get:
\begin{equation}
	\begin{gathered}
		b_M M^2- \left(\frac{J^2}{a_2} - b_{1M}  \right) 
  \Delta_1^2
 +
 \frac{J^2}{a_2}
 (
 2b_{2M}\Delta_1^2 M^2
 +
b_{12}\Delta_1^4
 )
 \approx 0,
		\\
		b_1 \Delta_1^2- \left(\frac{J^2}{a_2} - b_{1M}  \right) 
M^2
 +
 \frac{J^2}{a_2}
 (
 2b_{12}\Delta_1^2 M^2
 +
b_{2M}M^4
 )
 \approx 0.
	\end{gathered}
\label{eq:degpointappr}
\end{equation}
Solving these for $\Delta_1^2$ and $M^2$, we find:
\begin{align}
\Delta_1^2
&=
\frac{a_2^2 b_M}{3 J^2 b_{2M}} 
\frac{\left(\frac{J^2}{a_2}-b_{1M}\right)^2 - b_1b_M}
{\left(\frac{J^2}{a_2}-b_{1M}\right)\left(\frac{J^2}{a_2}-b_{1M}+\frac{b_{12}b_M}{b_{2M}}\right)} \notag\\
&\approx
\frac{4 a_2 (J-J_c)}{3 J_c
\left(
b_{2M}
\sqrt{\frac{b_1}{b_M}}
+
b_{12}
\right)
},
\\
M^2 &\approx \frac{4 a_2 (J-J_c)}{3 J_c
\left(
b_{12}
\sqrt{\frac{b_M}{b_1}}
+
b_{2M}
\right)
}
,
\notag\\
&\Delta_2 
\approx
-
\frac{4 (J-J_c)}{3
\left(
b_{12}
\sqrt{\frac{b_M}{b_1}}
+
b_{2M}
\sqrt{\frac{b_1}{b_M}}
\right)
},
\label{eq:OPscaling}
\end{align}
where $J_c = \sqrt{a_2(b_{1M}+\sqrt{b_1 b_M})}$.
In principle, these expression can be used to obtain corrections of the order $(J-J_c)^2$, but we have written only the leading order after the $\approx$ sign.

\subsection{First order transition temperature, latent heat and jump in specific heat}\label{appAnalytical}

We consider the case where $a_1=a_M\equiv a,\;b_1=b_2=b_M\equiv b,\;b_{12}=b_{1M}=b_{2M}=b'$. Due to symmetry of coefficients, $\Delta_1=\pm M$ will be a solution; we pick $\Delta_1=- M \equiv x$ without loss of generality. From the first two equations \eqref{eq:deglimitfull} we get:
\begin{align}
&a x + b x^3 - J x \Delta_2+ b' x^3 + b' \Delta_2^2 x= 0 \notag\\
&\qquad\qquad \implies x^2 = \frac{J \Delta_2-a- b'\Delta_2^2}{\tilde{b}},\label{eqx2}
\end{align}
where $\tilde{b} \equiv b+b'$. For this case one obtains a necessary condition $J \Delta_2>a>0$. Inserting this into equation for $\Delta_2$ one gets:
\begin{equation}
	\begin{gathered}
	a_2 \Delta_2 +b_2 \Delta_2^3 -J x^2 + 2 b'\Delta_2 x^2=
	\\
	\approx
	\frac{3 b' J \Delta_2^2-(J^2-a_2 \tilde{b}+2b'a)\Delta_2+Ja + O(\Delta_2^3)}{\tilde{b}} = 0.
	\end{gathered}
\end{equation}
Assuming weak first order transition, $\Delta_2^3$ term can be neglected and we obtain:
\begin{equation}
\Delta_2 = \frac{J^2 - a_2 \tilde{b}+2ab' \pm \sqrt{[J^2-a_2\tilde{b}+2b'a]^2-12 J^2ab'}}{6Jb'},
\end{equation}
where "+" sign is to be taken for the local minimum of free energy. For $J$ close to $J_c = \sqrt{a_2 \tilde{b}}$, one observes that the nontrivial solution exists only for $a < (J-J_c)^2/3b'$. In this regime, the condition $a<J\Delta_2\propto (J-J_c)$ is satisfied parametrically, note also that $ab'/J_c < (J-J_c)^2/3J_c\ll J-J_c$. The expression for $\Delta_2$ then simplifies to
\begin{equation}
\Delta_2 \approx \frac{J-J_c+\sqrt{(J-J_c)^2-3 ab'}}{3b'} 
\end{equation}
We can now also compute the free energy as a function of $a$ for this solution.
\begin{align}
    F_{\rm min} = a x^2+ \frac{a_2}{2} \Delta_2^2 - J x^2 \Delta_2+\frac{\tilde{b}}{2}x^4+ b' \Delta_2^2x^2+\frac{b_2}{4}\Delta_2^4
\end{align}
We drop the $O(\Delta_2^4)$ term and simplify \eqref{eqx2} by exploiting $J\Delta_2 \gg a,b'\Delta_2^2 (\sim (J-J_c)^2)  \implies x^2\approx J\Delta_2/\tilde{b}$:
\begin{equation}
F_{\rm min} \approx 
-\frac{((J-J_c)\Delta_2 - a- b' \Delta_2^2)J_c \Delta_2}{\tilde{b}}
\end{equation}
From here, one can find the critical value of $a$ from $F_{min}=0$
\begin{equation}
	a_{cr} = \frac{(J-J_c)^2}{4b'},
\end{equation}
evaluate the jumps in the order parameters at $T_c$
\begin{equation}
	\delta\Delta_2 = \frac{J-J_c}{2b'},\; \delta x = \sqrt{\frac{J_c(J-J_c)}{2b'\tilde{b}}} 
\end{equation}
and the expansion of the minimal free energy at $T\to T_c^-$:
\begin{align}
&F_{\rm min}(T-T_c) \notag\\
&\approx \frac{a_2 (J - J_c)}{2 b' J_c} (a-a_{cr}) - \frac{a_2}{2 J_c (J - J_c)} (a-a_{cr})^2 \notag\\
&=\frac{ a_2 (J - J_c)\alpha}{2 b' J_c} (T-T_{c}) - \frac{ a_2 \alpha^2}{2 J_c (J - J_c)} (T-T_{c})^2. \label{eqFminapp}
\end{align}
From \eqref{eqFminapp}, we can read off the latent heat and the specific heat jumps as $J\to J_c$
\begin{equation}
	\begin{gathered}
\Delta Q = T_c \Delta S \approx T_c \frac{a_2 (J - J_c) \alpha }{2 b' J_c},
\\
\frac{\Delta C_v}{T_c}  \approx \frac{ a_2 \alpha^2}{ J_c (J - J_c)},
\end{gathered}
\end{equation}
and find that the product of the latent heat and the specific heat jump approaches a constant as $J\to J_c$.
\begin{align}
    \frac{\Delta Q}{T_c} \frac{\Delta C_v }{T_c} \approx \frac{a_2^2 \alpha^3}{2J_c^2 b'}
\end{align}

\section{Analytical solution in presence of strain}\label{appRUS}

We now find the corrections to the above solution in the presence of small strain. The change in the sound velocities at the phase transition arises from the coupling of the strain to the order parameters captured by the free energy gathered from \eqref{eq:lin},\eqref{eq:quadr},\eqref{eqFOP}
\begin{align}
    F-F_N=&\frac{\bar{a_1}}{2} |\Delta_{B_{1u}}|^2 
    + \frac{\bar{a_2}}{2} |\Delta_{B_{2u}}|^2 
    + \frac{\bar{a_m}}{2} M_x^2 \notag \\
    &+ 2\lambda \varepsilon_{yz} \Re(\Delta_{B_{1u}} \Delta_{B_{2u}}^*) + J M_x \Im(\Delta_{B_{1u}} \Delta_{B_{2u}}^*) \notag \\
    &+ \frac{b}{4} (|\Delta_{B_{1u}}|^4 + |\Delta_{B_{2u}}|^4 +M_x^4) \notag \\
    &+ \frac{b'}{2} (|\Delta_{B_{1u}}|^2|\Delta_{B_{2u}}|^2 +|\Delta_{B_{1u}}|^2|M_x^2 + |\Delta_{B_{2u}}|^2|M_x^2)  \notag \\
    &+ \frac{b''}{2} ((\Delta_{B_{1u}}^*)^2 \Delta_{B_{2u}}^2 + (\Delta_{B_{2u}}^*)^2 \Delta_{B_{1u}}^2) \label{eqFbar}
\end{align}
where $\bar{a_i}=a_i + \sum_\alpha \lambda^{(1)}_{\alpha,i} \varepsilon_{\alpha\alpha} + \sum_\alpha \lambda^{(2)}_{\alpha,i}  \varepsilon_{\alpha\alpha}^2 + 2 \sum_{\alpha\beta} \lambda^{(2)}_{\alpha\alpha,\beta\beta,i} \varepsilon_{\alpha\alpha} \varepsilon_{\beta\beta} + \sum_{\alpha\beta} \lambda^{(2)}_{\alpha\beta,i} \varepsilon_{\alpha\beta}^2$  for $i\in (1,2,m)$. The linear coupling to the shear strain induces a correction to the minimal values of the order parameter found in the previous section
\begin{align}
    \Delta_{B_{1u}}=-M=x, \, \Delta_{B_{2u}} =\Delta_2' +i\Delta_2'', \, \Delta_2'\approx -\frac{2\lambda \varepsilon_{yz} x }{a_2}
\end{align}
with $x \approx \sqrt{\frac{J\Delta_2''}{\tilde{b}}}$ and $\Delta_2'' \approx  \frac{J-J_c+\sqrt{(J-J_c)^2-3 ab'}}{3b'}$ as before.
With these values inserted in \eqref{eqFbar}, the minimum of the free energy just below $T_c$ takes the form in \eqref{eqFmin}, from which the jumps in the shear moduli can be extracted by taking the appropriate derivatives in presence of the order parameters
\begin{align}
    \Delta c_{44} =\frac{\partial^2 (F-F_N)}{\partial \varepsilon_{yz}^2}\bigg|_{\{\varepsilon\}=0} \notag \\
    \Delta c_{55} =\frac{\partial^2 (F-F_N)}{\partial \varepsilon_{xz}^2}\bigg|_{\{\varepsilon\}=0} \notag \\
    \Delta c_{66} =\frac{\partial^2 (F-F_N)}{\partial \varepsilon_{xy}^2}\bigg|_{\{\varepsilon\}=0} 
\end{align}

\end{document}